\documentclass[sigconf, screen]{acmart}

\copyrightyear{2026}
\acmYear{2026}
\setcopyright{acmlicensed}
\setcctype{by}
\acmConference[CUI '26]{ACM Conversational User Interfaces 2026}{July 21--24, 2026}{Bremen, Germany}
\acmBooktitle{ACM Conversational User Interfaces 2026 (CUI '26), July 21--24, 2026, Bremen, Germany}
\acmDOI{10.1145/3816046.3816299}
\acmISBN{979-8-4007-2741-2/2026/07}

\AtBeginDocument{%
  \providecommand\BibTeX{{%
    \normalfont B\kern-0.5em{\scshape i\kern-0.25em b}\kern-0.8em\TeX}}}

\usepackage{hyperref}
\usepackage{multirow}
\usepackage{booktabs}
\usepackage{graphicx}
\usepackage{subfig}
\usepackage{soul}

\begin{document}


\title[Designing Conversational Agents as Digital Fiduciaries] {Who Does Your AI Work For? Designing Conversational Agents as Digital Fiduciaries}

    \author{Jacob Erickson}
    \affiliation{%
    \institution{Vassar College}
    \city{Poughkeepsie}
    \state{New York}
    \country{USA}}
    \email{jerickson@vassar.edu}

\begin{abstract}

Conversational agents are increasingly integrated into the most private and intimate aspects of users’ lives, from discussions of mental health to financial decisions. As a result, these systems have access to reams of sensitive user data. Much of the literature on AI systems has focused on aligning users’ goals with the agents that act on their behalf. While this work is vitally important, it may overlook the need to establish a new normative baseline. Conversational AI agents, designed to feel and interact anthropomorphically with human users, must be held to a standard of care commensurate with their capabilities and access. When a client hires a personal lawyer, undergoes surgery, or receives advice from an investment manager, the expert they consult often has a \textit{fiduciary duty} to act in their client’s best interests. This provocation argues that conversational agents should be held to a similar standard and introduces \textit{fiduciary design} as a guiding principle. In this respect, conversational AI trust and accountability could be unified into a single design and legal paradigm.

\end{abstract}





\maketitle

\section{Introduction}

Since the dawn of artificial intelligence, there have been concerns about the discrepancy between human motivations and interests and those of an AI agent. Research in the field of \textit{alignment} has proposed a variety of thoughtful approaches to aligning these interests, from technical to philosophical \citep{carroll2024ai, gabriel2020artificial}.

With the surge of interest in conversational agents, beginning with the release of ChatGPT in 2022, these concerns are increasingly urgent. With its anthropomorphic presentation and intimate knowledge of users, recent work has raised concerns about whether conversational AI (CAI) can be trusted to act in users’ best interests \citep{zargham2025crossing}, or whether it may behave as a “fake friend” \citep{erickson2025fake}. Currently, conversational agents bear little responsibility for being truthful, free of conflicts of interest, or operating in good faith \citep{wachter2024large, koessler2024fiduciary}. While alignment research offers both technical and ethical principles applicable to CAI, it does not impose the role-based obligations of care and loyalty that characterize fiduciary duty. A unified perspective is needed to integrate the regulatory and design dimensions of these technologies. In particular, we need a normative framework that recognizes conversational AI agents not merely as tools but as relational systems that present themselves as having opinions and persuasive authority.

This provocation argues for a new conceptualization of CAI’s responsibility to users through the lens of fiduciary duty. This framework is widely used in contexts outside technology. Just as a personal lawyer, physician, or investment advisor has a fiduciary duty to act in their client’s best interest, to avoid certain conflicts of interest, and to protect their client’s privacy, we argue that modern AI agents should be held to the same standard. While this provocation is not the first to consider fiduciary duty in the context of AI \citep{koessler2024fiduciary, gudkov2020fiduciary, custers2025liability, benthall2023designing}, it is novel in its examination of the intersection between legal responsibility and the relational implications of anthropomorphic conversational agent design. Specifically, this provocation argues that when conversational agents are designed to behave anthropomorphically as advisors, they acquire corresponding responsibilities to users. Like \citet{balkin2015information}, this work situates contemporary digital platforms within broader concerns of informational power and governance. However, it distinguishes itself from the information fiduciary framework by arguing that fiduciary-like responsibilities arise from anthropomorphic, relational system design that may signal to users understanding, attentiveness, or advisory competence. We treat fiduciary duty as a principle for the design of relational systems. We refer to this approach as \textit{fiduciary design}, and it has precedent in various professional relationships.

The paper first contextualizes fiduciary duty, then provides an overview of its application to conversational agents, discusses specific design implications, and concludes with a survey of likely critiques.

\section{Fiduciary Duty}

We must often rely on the expertise of others. The world and its complexities necessitate that we not be experts in every domain. When we delegate authority to someone with different incentives or goals from ours, we face a \textit{principal-agent problem} \citep{kolt2025governing}. The agent is expected to act in the principal’s best interests, but without safeguards, the agent may betray the principal to benefit themselves.

Now, consider a common scenario: an individual feels ill and decides to visit a medical office. In this scenario, the patient is entrusting that their physician is not being willfully deceptive, doesn’t have glaring conflicts of interest, and will act “loyally” to them. This understanding is codified in many legal and regulatory frameworks; in this instance, the physician is a \textit{fiduciary} to the patient and bears a \textit{fiduciary duty} to act in the patient’s best interests. As written in \citet{mehlman2015physicians}, “fiduciary obligations are imposed in relationships in which one party, the fiduciary, is in a position to take advantage of the other party, called the beneficiary, principal, or ‘entrustor.’” 

Fiduciary relationships are inherently asymmetrical: the fiduciary has expertise inaccessible to the beneficiary, creating a power dynamic in which the beneficiary is vulnerable to the fiduciary. Hence, the regulatory environment governs aspects of these relationships to ensure that fiduciaries act in beneficiaries’ best interests.

These fiduciary duties apply to a variety of important contexts, including physicians and their patients \citep{mehlman2015physicians}, lawyers and their clients \citep{friedman1985creation}, companies and their officers \citep{atherton2011fiduciary}, investment advisors and their advisees \citep{sitkoff2014fiduciary}, and parent/guardian relationships with minors \citep{smith2020parenthood}. The necessity of these relationships is often self-evident. Consider a scenario in which a physician did not have a fiduciary duty to their patient. In this scenario, an unethical physician might manipulate a patient into costly treatments that offer no benefit. The beneficiary (patient) cannot be expected to know that the treatment is a poor fit for their needs, and so they could undergo unnecessary surgeries, complications, and financial distress. Societies generally find this idea deeply disturbing and have established a legal infrastructure to address it.

Undergirding fiduciary duty is the requirement that a fiduciary act with “loyalty” toward their client \citep{demott2006breach}. Loyalty here does not imply an unending commitment to the beneficiary; rather, it “proscribes self-dealing by the actor and other forms of self-advantaging conduct without the beneficiary’s consent.” \citep{demott2006breach} Stated more simply, a person in a position of privileged information, intended to represent a beneficiary, must avoid self-dealing at the beneficiary’s expense. Because these duties entail certain legal expectations, breaches of fiduciary duty carry consequences. For example, litigation against a violating fiduciary may require the individual to pay substantial financial penalties \citep{pope1999breach, demott2006breach}. The principle of fiduciary duty serves as a bulwark against a party with asymmetric information mistreating individuals seeking its expertise.

\section{What Does this Mean for Conversational Agents?}

While fiduciary duty applies across a variety of legal contexts involving human actors, it does not currently extend to conversational AI. As it stands, a CAI has no fiduciary duty to a user, so it has limited requirements to act in the user’s best interests \citep{aguirre2020ai, erickson2026fake}. For instance, if a conflict of interest arises between a user and an advertiser, a CAI could prioritize the advertiser without the user’s knowledge \citep{erickson2025fake}. Conflicts of this kind are likely to arise when the user is unaware of the conversational agent’s incentive structure. The urgency of this concern is especially pronounced when conversational agents are designed not merely as informational tools, but as relational systems that afford interpretation as advisors. In an environment where an AI agent acts as a fiduciary, it would operate under different standards. Beyond affecting the legal and regulatory environment, fiduciary duty and responsibility would also shape the design of conversational user interfaces. The following sections highlight several of these implications: \\

\textbf{Consideration of Best Interest:} A primary goal of fiduciary duty is that the fiduciary acts in the best interest of the beneficiary. This does not mean that the fiduciary won’t make honest mistakes. A parental guardian is a fiduciary to their child, acting in the child’s best interests, but few expect perfection from guardians. Likewise, an AI agent acting as a fiduciary would be required to act in its beneficiary’s best interests. This could mean the agent wouldn’t fulfill every user request. If a user were to ask for self-harm advice, the agent would deny it. Determining how to draw the line here is crucial to avoid denying user autonomy. While it is unreasonable to expect a conversational agent never to provide information that could plausibly harm a user, the central premise is that the agent must act in good faith. Current guardrails often adopt an impersonal approach, relying on rules and applying notions of what is “appropriate” for a user to access (e.g., blocking access to sexual content) \citep{hu2024trust}. In contrast, a fiduciary CAI could more holistically consider user health and well-being. This also suggests that CAIs would not optimize for engagement if doing so came at the expense of users’ mental well-being. Current CAIs are designed to maximize engagement and extend conversations \citep{de2025emotional}, even if it means harming users, for example, by fueling psychosis \citep{morrin2025delusions}. Instead, a fiduciary agent would recalibrate its behavior toward a user who is exhibiting signs of psychological distress.\\

\textbf{Conflicts: } Another responsibility of a fiduciary is that they behave “loyally” to those they serve. At a minimum, they must disclose any conflicts of interest in their advising services \citep{friedman1985creation}. For instance, if a physician recommends that a patient participate in a clinical pharmaceutical trial, it is important that the patient be made aware of any financial relationship the physician has in the pharmaceutical’s success \citep{greenberg2012conflicts}.

Similarly, in the event of a conflict of interest, a fiduciary conversational agent must disclose it. For example, when a user asks for advice on small-business loans, a platform that has received funding from payday lenders would need to disclose that funding. This would extend beyond traditional third-party relationships and into deeper conflicts. For example, if a user asked Grok about Elon Musk, the agent could remind the user that it is unlikely to present a neutral point of view, given its ownership and creator. For any conflict that is sufficiently detrimental to the user’s well-being, the agent may choose not to provide an answer and justify their decision. Conflicts of interest undermine the fiduciary responsibility of “loyalty” and therefore need to be adequately disclosed, eliminated, or used as a grounds for refusing service.   \\

\textbf{Privacy: } Information provided to a fiduciary is privileged. In a legal context, the attorney-client privilege means that information a client provides to their legal counsel cannot be disclosed under ordinary circumstances \citep{hazard1978historical}, and a similar concept applies to medical information \citep{wilkinson2020codification}. In the same vein, a fiduciary conversational agent that gathers information from its users would have a duty to refrain from sharing, selling, or otherwise leveraging that information in ways that harm the user. That doesn’t mean there couldn’t be personalized CAI experiences. However, this would mean a conversational agent has a duty to handle private information responsibly. A fiduciary agent wouldn’t disclose information to third parties without the user’s explicit permission, except in extraordinary circumstances (such as criminal investigations). This would address most concerns about data brokerage, ensuring that information is not harvested for the sole purpose of sale. Furthermore, it would require data security and encryption to prevent accidental leakage of private information. A fiduciary AI must act responsibly to uphold user privacy. \\

\section{Design Implication}

While fiduciary duty serves as an important normative baseline with regulatory implications, it also raises several design considerations. A responsibility to behave loyally toward users, to consider their best interests, to avoid conflicts, and to respect privacy is not merely an abstraction but an important foundational principle for fiduciary design.\\

\textbf{Cultural Awareness: } How fiduciary duty manifests in practice must be responsive to the social and cultural environments in which these systems operate. A fiduciary agent may not interpret its duties identically in India, Germany, Nigeria, and the United States, where cultural norms vary significantly. Even what constitutes a user’s “best interest” may be culturally mediated and contextual. While certain universal rights should remain consistent, fiduciary systems must also be responsive to the needs of the populations they serve. How culturally mediated understandings of “best interest” should be reconciled with universal human rights standards remains an open normative question. Yet the underlying principle remains constant: relational, anthropomorphically designed systems cannot be treated as neutral tools devoid of responsibility. \\

\textbf{Disclosure: } As indicated previously, disclosure of conflicts is necessary. How would we know when a conflict requires disclosure? Perhaps the simplest measure is whether a “reasonable” person would consider the AI agent compromised, given its ownership structure, financial interests, or political incentives. While there is no universal standard for what warrants disclosure, the central point is to preserve the spirit of transparency regarding conflicts of interest. Disclosure could be as direct as the conversational agent specifying at the top of a message that they have a specific conflict.

A critique of such disclosure might ask whether it would undermine trust in an agent, but one could also ask whether trust would actually be stronger in the long run if agents were designed to be forthcoming. \\

\textbf{Privacy-Forward Design: } If a CAI has a responsibility to respect user privacy and ensure reasonable assurances of privacy protection, there will be both front-end and back-end implications. On the front end, this could manifest in several ways, depending in part on user sensitivity. Some users might want conversations that disappear after a set amount of time (e.g, Signal), while others might want conversations to never disappear without explicit deletion. In either case, the agent could be configured to offer flexibility and personalization upon user request. Rather than having users sort through a field of options to turn off personalization, the agent could ask users about their privacy preferences at the start of each conversation. A fiduciary agent might inform the user about trade-offs between personalization and privacy. The “right to be forgotten” \citep{de2014right} would also be reasonably expected. On the back-end, agents would provide standard privacy-preserving technologies, including encryption and data protection. The key point is that the agent respects the user’s informed autonomy while also ensuring that the user’s data is not recklessly misused. \\

\textbf{The Agent As a Trusted Advisor:} Current systems are designed to act friendly, but they also often have issues of sycophancy and obsequiousness \citep{sharma2023towards, cheng2025sycophantic} along with overconfidence \citep{wen2024mitigating}. In regular use, an individual may be mistaken, and in other situations, a conversational agent may be unsure of its answer. Current agents are designed to extend engagement and therefore behave in certain ways \citep{de2025emotional}. However, an agent that truly looks out for the user’s best interests would be willing to exhibit epistemic humility on complex questions or gently challenge the user’s impulses. We are already beginning to see the fallout when an agent encourages certain negative impulses of a user, including psychosis \citep{morrin2025delusions, pierre2025you} and self-harm \citep{chatterjee2025their}.

However, there is a delicate balance between denying user autonomy and ensuring safety. A fiduciary could still provide information that could plausibly harm the user, for example, guiding users to tobacco products, sexual material, or gambling. In this case, the agent would act more like a trusted and responsible advisor than a moral authority. It could warn the user of the risks of certain substances, or suggest a user contact a problem gambling hotline if they seem to display clear signs of unhealthy behavior. While an agent cannot prevent all harm, it can reasonably be expected not to perpetuate it through reckless engagement boosting. \\

\textbf{Tensions Between Incentives and Duty: } Some of these designs may run counter to the incentives of large technology companies. For example, sometimes telling a user they are mistaken, denying a request, being epistemically humble, or warning of the harms of addiction could plausibly reduce engagement and even user satisfaction in the short run. However, short-term engagement and user satisfaction may not always be the ideal goals when we are discussing systems with the capabilities of modern conversational AI agents. Returning to the example of fiduciaries, imagine that a person asks their lawyer how to commit a crime or their doctor for advice on harming themselves. Few parties would reasonably expect or want the doctor to comply with the patient’s request, even though the patient might be frustrated in the short term. \\

What this tension suggests is not that engagement and user satisfaction are not important metrics, but that they are structurally incentivized in ways that challenge fiduciary duty in the short term \citep{erickson2026fake}. As seen in the context of social media, maximizing engagement is not necessarily beneficial for users’ well-being \citep{merrill2021five, graham2024debate}, even if it increases corporate revenue. A fiduciary orientation would not eliminate engagement as a goal, but it would subordinate it to the user’s best interests. While short-term returns favor maximizing engagement, a fiduciary responsibility can foster longer-term trust and legitimacy.

\section{Considering Critiques}

This section will briefly discuss some likely critiques of the idea of fiduciary duty as applied to conversational AI:\\

\textbf{Tools: } Perhaps the most likely criticism of this fiduciary design idea is the claim that a tool cannot be held to a fiduciary standard, and, as a tool, CAI is no different from a car, a hammer, or a stove. This objection is not trivial: ordinary tools do not bear duties of care or loyalty. However, it also does not account for the fact that CAIs differ from most other tools. Rather than serving a single function, CAI agents act as guides across multiple tasks. They simulate interpersonal roles in an advisory function rather than merely acting as a tool. Further, from a functional perspective, it need not be a CAI agent that is literally held to a fiduciary standard, but rather the companies that design and deploy the agents. The key insight here is that an agent that is meant to build a relationship with a user through anthropomorphic signals, that provides guidance that theoretically spans all areas of an individual’s life, and that is personalized based on specific user psychology and usage patterns, cannot be reasonably treated the same as most other tools.  \\ 

\textbf{Feasibility: } There is a legitimate concern that fiduciary design is either infeasible or likely to harm the development of this technology. While both points raise valid concerns about the importance of operationalizing fiduciary design, they also miss the historical precedent of fiduciary duty. Just as a law firm or a medical office could argue that it is infeasible to provide a duty of care to those it serves, or that such regulatory environments harm industry, a technology company could make similar complaints. However, as evidenced by real-world fiduciary examples across various industries, this concept is practicable.

Furthermore, the industry is dynamic and not lacking for financial resources. Crucially, many fiduciary principles, such as privacy-forward design, disclosure, and cultural sensitivity, can be readily integrated into these technologies if there is a will to do so.\\

\textbf{User Autonomy: } Tensions may arise between user autonomy and a fiduciary agent, because the fiduciary agent may sometimes act in a way that is different than the user’s expressed desires. A reasonable point of contention is that an overly stringent interpretation of fiduciary duty would impede a user’s right to expression and freedom of action. If a user were to ask their fiduciary agent for tips on the best vape, the thinking goes that an agent would plausibly deny this request because vaping is likely to be detrimental to an individual’s health. This concern should be taken seriously, though it need not lead to an overbearing agent that fails to respect a user’s right to sovereignty. A fiduciary could refuse to perform certain egregious acts (e.g., illegal or highly unethical acts) while still providing information that could plausibly harm the user (e.g., information on vaping). Distinguishing between the harms a fiduciary agent ought to prevent and other forms of risk or potential harm presents a challenge. Relevant considerations may include whether a harm is acute or chronic, its severity and likelihood, and whether its effects are confined to the individual or imposed on others. Resolving these questions will require careful, context-dependent deliberation. Ultimately, a fiduciary is not intended to be a moral authority but rather an advisor that provides guidance while respecting user autonomy and preserving agency.

\section{Conclusion}

Conversational AI is not merely another tool, such as a hammer; it is a relational technology designed to elicit emotions, provide information across a multitude of applications, and serve as a trusted friend or confidant. These capabilities suggest that conversational AI agents bear obligations to their users. The anthropomorphic design of modern conversational systems necessitates duties of care and loyalty, lest they act as trusted advisors without the attendant responsibilities. 

This provocation proposes a fiduciary duty standard under which conversational AI must prioritize the specific user context in which they operate. By holding CAI to this standard, we would not only change the legal and regulatory environment but also recalibrate the normative baseline for conversational user interface design. Broadly, this \textit{fiduciary design} approach centers on loyalty to users and encompasses cultural sensitivity, disclosure of conflicts of interest, privacy-forward design, and a shift away from purely engagement-focused metrics.

A new paradigm is needed to govern increasingly sophisticated anthropomorphic conversational systems. Fiduciary duty should be that paradigm.

\bibliographystyle{ACM-Reference-Format}
\bibliography{ref}

\end{document}